\documentclass[preprint]{aastex6}

\usepackage{graphics,graphicx}

\usepackage{natbib}
\usepackage{amsmath}

\usepackage{color}
\definecolor{red}{rgb}{1.,0.0,0.}

\begin{document}

\title{The Orbit and Mass of the Cepheid AW Per\footnote{Based on    observations with the NASA/ESA {\it Hubble Space Telescope} obtained at the Space Telescope Science Institute, which is operated by the Association of Universities for Research in Astronomy, Inc., under NASA contract NAS5-26555. }}



\author{Nancy Remage Evans}
\affil{Smithsonian Astrophysical Observatory,
MS 4, 60 Garden St., Cambridge, MA 02138; nevans@cfa.harvard.edu}

\author{Alexandre Gallenne}
\affiliation{Instituto de Astrof\'isica, Departamento de Ciencias Físicas, Facultad de Ciencias Exactas, Universidad Andr\'es Bello, Fern\'andez Concha 700, Las Condes, Santiago, Chile and French-Chilean Laboratory for Astronomy, IRL 3386, CNRS, Casilla 36-D, Santiago, Chile}

\author{Pierre Kervella}
 \affil{LESIA, Observatoire de Paris, Universit\'e PSL, CNRS, Sorbonne Universit\'e, Universit\'e de Paris, 5 Place Jules Janssen, 92195 Meudon, France}

\author{Antoine M\'erand}
 \affil{European Southern Observatory, Karl-Schwarzschild-Str. 2, 85748 Garching, Germany}

 \author{John Monnier} 
\affiliation{University of Michigan, Department of Astronomy,
1085 S. University, Ann Arbor, MI 48109}

\author{Richard I Anderson}
\affil{Institute of Physics, \'Ecole Polytechnique F\'ed\'erale de Lausanne (EPFL), Observatoire de Sauverny, 1290 Versoix, Switzerland}

\author{H. Moritz G\"unther}
\affil{Massachusetts Institute of Technology, Kavli Institute for Astrophysics and
Space Research, 77 Massachusetts Ave, NE83-569, Cambridge MA 02139, USA}

\author{Charles Proffitt}
\affiliation{Space Telescope Science Institute, 3700 San Martin Drive, Baltimore, MD 21218}

\author{Elaine M. Winston}
\affil{Smithsonian Astrophysical Observatory,
MS 4, 60 Garden St., Cambridge, MA 02138; nevans@cfa.harvard.edu}

\author{Grzegorz  Pietrzynski}
\affiliation{Departamento de Astronomia, Universidad de Concepcion, Casilla160-C, Concepcion, Chile}

\author{Wolfgang Gieren} 
\affiliation{Departamento de Astronomia, Universidad de Concepcion, Casilla160-C, Concepcion, Chile}

\author{Joanna Kuraszkiewicz}
\affil{Smithsonian Astrophysical Observatory,
MS 67, 60 Garden St., Cambridge, MA 02138; nevans@cfa.harvard.edu}

\author{Narsireddy Anugu}
\affil{CHARA Array of the Georgia State University, Mount Wilson, CA, 91023}

\author{Rachael M. Roettenbacher}
\affil{University of Michigan, Department of Astronomy,
1085 S. University, Ann Arbor, MI 48109}

\author{Cyprien Lanthermann}
\affil{CHARA Array of the Georgia State University, Mount Wilson, CA, 91023}

\author{Mayra Gutierrez}
\affil{CHARA Array of the Georgia State University, Mount Wilson, CA, 91023}
\affil{University of Michigan, Department of Astronomy,
1085 S. University, Ann Arbor, MI 48109}

\author{Gail Schaefer}
\affil{CHARA Array of the Georgia State University, Mount Wilson, CA, 91023}

\author{Benjamin R. Setterholm}
\affil{Max-Planck-Institut f\"ur Astronomie, K\"onigstuhl 17, 69117 Heidelberg
Germany}

\author{Noura Ibrahim}
\affil{University of Michigan, Department of Astronomy,
1085 S. University, Ann Arbor, MI 48109}

\author{Stefan Kraus}
\affil{University of Exeter, School of Physics and Astronomy, Astrophysics Group,
  Stocker Road, Exeter, EX4 4QL, UK }

\begin{abstract}
  The Cepheid AW Per is a component in a multiple system with a
  long period orbit.  The radial
  velocities of Griffin (2016) cover the 38 year orbit well.  An extensive
  program of interferometry with the CHARA array is reported here, from
  which the long period
  orbit is determined.  In addition, a {\it Hubble Space Telescope}
  high resolution spectrum in the ultraviolet demonstrates that the
  companion is itself a binary with nearly equal mass components.  These
  data combined with a distance from {\it Gaia} provide a mass of the Cepheid
  (primary) of M$_1$ = 6.79 $\pm$ 0.85 $M_\odot$.  The combined mass of the
  secondary is  M$_S$ = 8.79 $\pm$ 0.50 $M_\odot$. The accuracy of the mass will be
  improved after the fourth Gaia data release expected in approximately two years.  

\end{abstract}




\section{Introduction}\label{intro}

The quest for masses of Cepheid variable stars starts with the identification of binary or multiple 
systems of which they are members.  In the case of the Cepheid AW Per the possibility of a
companion was suggested both by photometry and radial velocities.  Oosterhoff (1960) suggested a 
hot companion could explain the relatively small photometric  amplitude in the ultraviolet.  
Miller and Preston (1964) 
obtained spectra and identified a composite spectrum made up of the Cepheid and hot companion.
The resultant spectrum had a combined   Ca II H (Cepheid) plus  H$\epsilon$ (companion) feature
which is deeper than the   Ca II K line from the Cepheid alone. 
Lloyd Evans (1968) was the first to find orbital motion in the system by comparing the 
velocities from Miller and Preston with earlier velocities from Joy (1937). 

Not surprisingly, many people over many seasons provided velocities for an orbit, which ultimately 
proved to be about  40 years (Welch and Evans 1989; Evans, Vinko, and Wahlgren 2000; and references
therein).  The 
difficulty of combining velocities from many instruments over a period this long was overcome by 
Griffin (2016) in a series of observations each year from 1976 to 2016 with the Cambridge
radial velocity spectrometer.  This comprehensive data string has produced a definitive
spectroscopic orbit.  The 
Griffin paper also contains an extensive summary of all observations of AW Per (of which only 
a summary is provided here).

Many developments since the discovery of a companion have helped toward the ultimate goal of 
deriving a Cepheid mass.  Oosterhoff, and Miller and Preston correctly identified the companion 
as a late B star.  Satellite ultraviolet spectroscopy provides a  spectrum of the
hot companion which is essentially uncontaminated by the Cepheid for wavelengths shorter than 
1700 \AA.   An {\it International Ultraviolet Explorer (IUE)} low resolution spectrum has 
been discussed several times (B\"ohm-Vitense and Proffitt 1985; Evans 1989; Welch and Evans 1989;
and Massa and Evans 2008) partly because of revised software (Evans 1994).  The temperature of
the companion is revisited below based directly on the comparison with Kurucz ATLAS9 atmospheres.
 Both the 
temperature inferred for the companion and the orbit have changed somewhat throughout various 
analyses.  However, the conclusion originally put forward by Welch and Evans that the mass function from 
the orbit and a reasonable mass estimate for the Cepheid require a companion mass larger than 
that of a single main sequence companion
 remains.  Thus, the most direct interpretation 
is that the companion is itself a binary.  This question is also discussed further below.

To derive a mass for the Cepheid, measured values of the inclination and the
angular separation in  arcseconds 
need to be added to the parameters of the Cepheid orbit.  Two new approaches have added these.  
First, Massa and Evans (2008) observed the AW Per system with the {\it Hubble Space Telescope} Space 
Telescope Imaging Spectrograph ({\it HST}/STIS), observing the system with three satellite roll 
angles.  This allowed them to measure the separation and position angle.  This observation is 
discussed in Section~\ref{hst.stis} (with a small update).

The second approach came from interferometry with the CHARA Array.   Data  
analyzed using the CANDID software package  resolved the system (Gallenne et al. 2015).
  

The Cepheid AW Per pulsates in the fundamental mode with a period of 6.4 d. 
In addition to the three stars making up the primary and the binary secondary,
Kervella et al. (2019b) have identified an additional probable bound companion 
at an apparent separation of 8400 au, which has a spectral type of approximately 
K3.5 V.

This paper contains  sections on the ultraviolet spectra, both the {\it HST} high
resolution spectra and an {\it IUE} low resolution spectrum, followed by CHARA
interferometry,  fitting  the orbit, including the Cepheid mass, and discussion
of the results.



 


\section{Ultraviolet Spectra}\label{uvspec}

\subsection{High Resolution}\label{hires}

In order to explore the companion further,  high resolution echelle ultraviolet
STIS spectra were obtained with  {\it HST} where the hot companion AW Per B
dominates completely.  The 
 observations have a central wavelength of 1416 \AA\/ and cover approximately
 1325 to 1500 \AA.  They were obtained 
over six orbits  between JD   2457301.9927  and 2457303.1391 for a total
exposure time of 15810 sec.

The echelle orders were combined using an appropriate blaze function, 
producing a  1D spectrum.  Interstellar lines (narrower than stellar lines)
were removed by  interpolation.  Details of the reduction are discussed for the 
V350 Sgr system (Evans, et al. 2018), including coadding the spectra from the
six orbits into a single spectrum.

These spectra were compared with that of $\xi$ Oct = HD 215573 using a 
spectrum from the Astral Hot Stars project 
(Ayres 2010\footnote{  https://casa.colorado.edu/\textasciitilde ayres/ASTRAL/}), 
a collection of high signal to 
noise, high resolution STIS spectra carefully processed. 
$\xi$ Oct is listed in this atlas as a B6 IV sharp-lined ``normal'' B star 
(V = 5.31 mag; B-V = -0.12 mag). It is `slowly pulsating B star', a class which 
has non-radial g-mode pulsations (albeit with small magnitude variations).
 One reason $\xi$ Oct was selected is that its parameters including temperature have been well-determined by Fitzpatrick and Massa (2005).

\begin{figure}
\plottwo{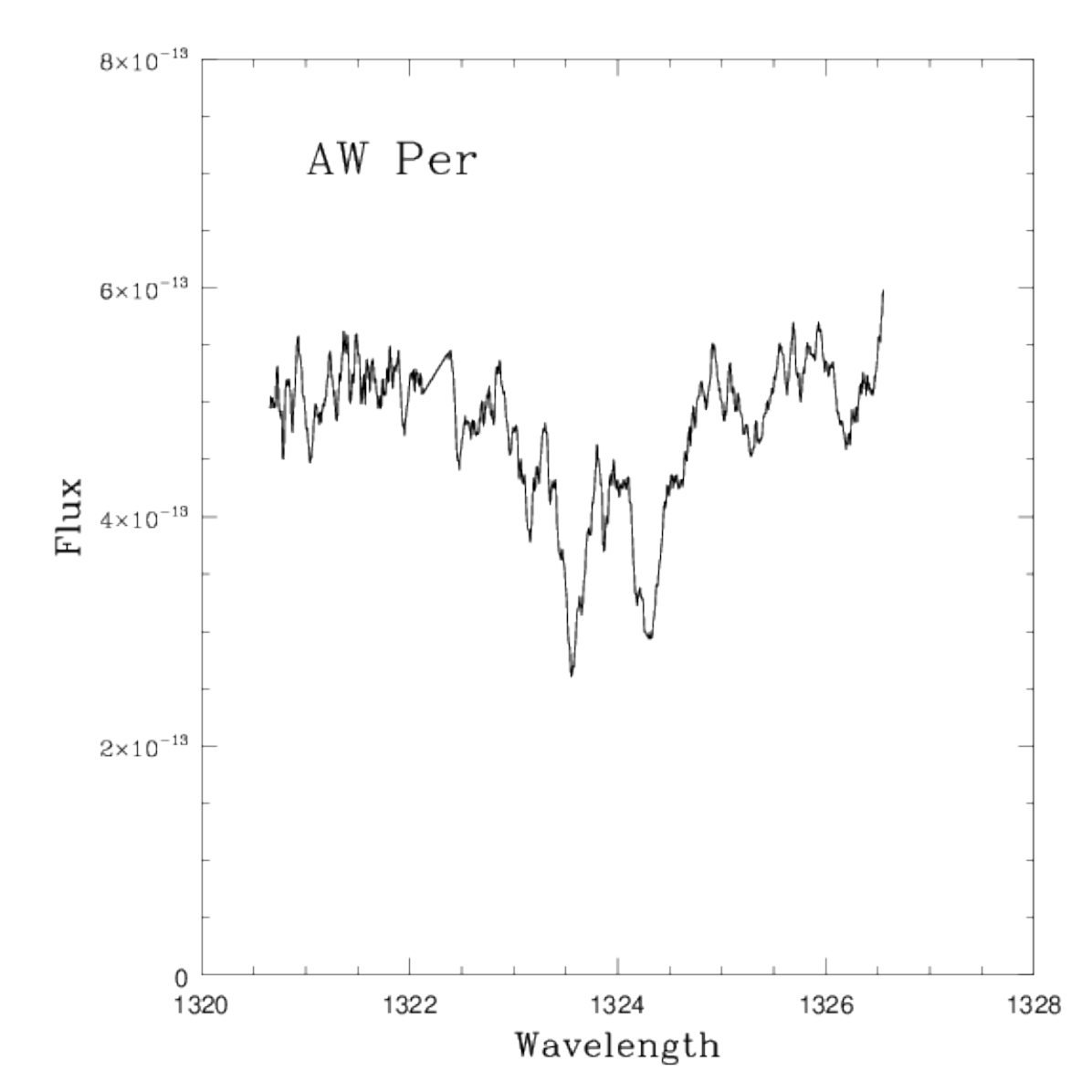}{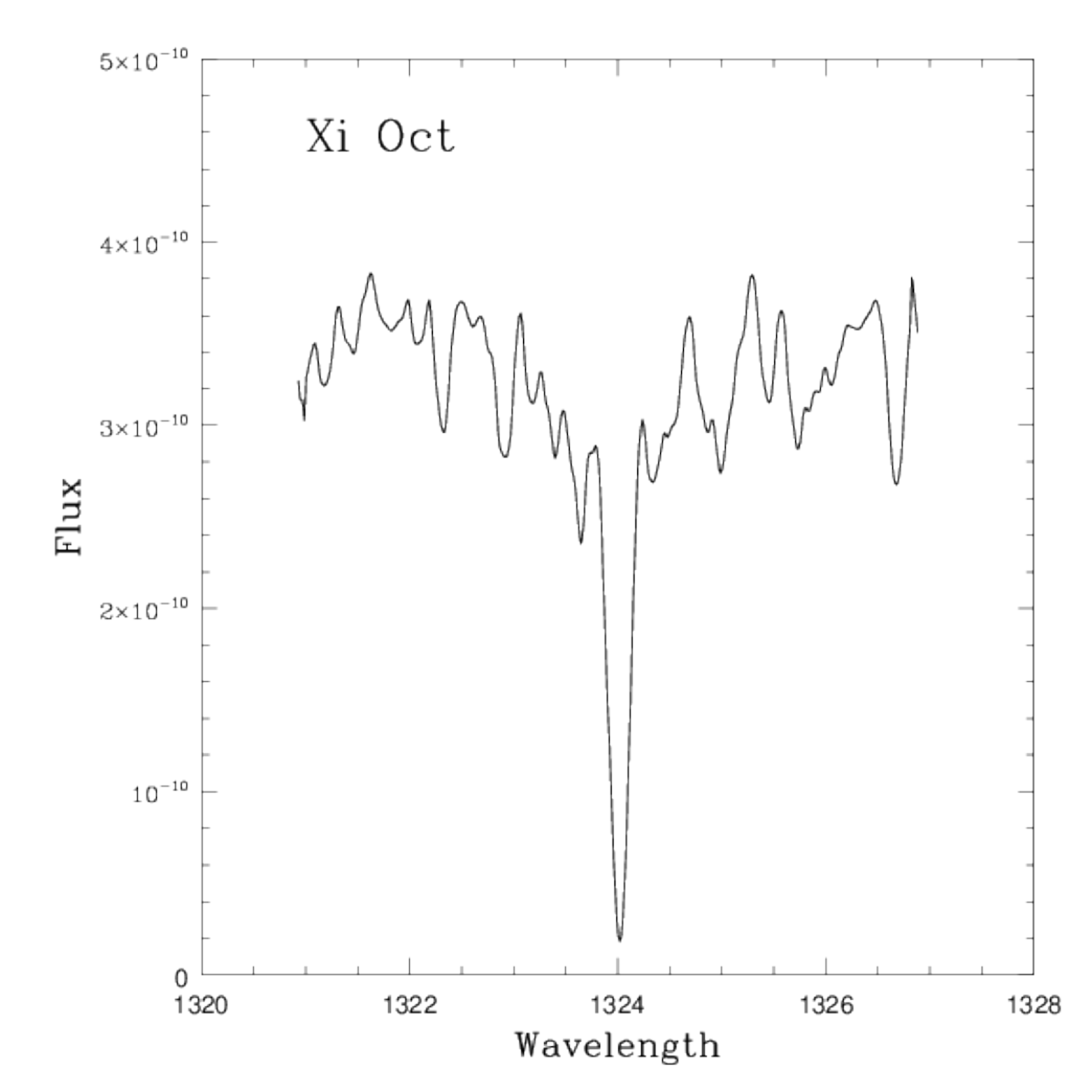}
\caption{Left: The spectrum of the hot companion
  AW Per B in the region of the C II line 1324\AA. Right: 
The spectrum of the B6 IV star $\xi$ Oct in the same region. The AW Per spectrum  
 has a 10 point boxcar smooth.   Wavelength is in \AA; flux
 is in ergs cm$^{-2}$ s$^{-1}$ \AA$^{-1}$.   
\label{lines}}
\end{figure}

The first inspection of the spectra of AW Per and $\xi$ Oct presents a striking contrast
(Fig.~\ref{lines}).  The C II line at 1324~\AA\/ has  the profile of a deep line
 in $\xi$ Oct.  
AW Per  has two separated but fairly narrow lines. That is both components of the 
binary companion Ba and Bb have been detected with a substantial velocity difference.




In Fig~\ref{lines}, the depths of the red and blue components of AW Per B are nearly 
equal, implying that the two stars Ba and Bb are similar in temperature 
and luminosity.
As an exploration of the properties of the components of the binary companion 
the spectrum of $\xi$ Oct has been used.  The temperature of   $\xi$ Oct has been
determined to be 14347 $\pm$ 138 K by Fitzpatrick and Massa from the energy distribution,
including  {\it IUE} spectra.  This is comparable to the temperature of AW Per found in
Section~\ref{lores}. 
An important part of the 
current study is to determine the properties of both stars in the binary companion
as far as possible.
In this exploration the temperatures of AW Per B and $\xi$ Oct are similar enough  for
an initial examination.

\begin{deluxetable}{llrrrr}
\tablecaption{Companion Velocities\label{vr}}
\tablewidth{0pt}
\tablehead{
 \colhead{$\lambda$} &  \colhead{$\lambda$}  & \colhead{V$_B$}  &  \colhead{$\pm$}
  & \colhead{V$_R$}
  & \colhead{$\pm$}     \\
 \colhead{Start} &  \colhead{End}  & \colhead{}  &  \colhead{}  & \colhead{}   & \colhead{}
 \\
 \colhead{\AA} &  \colhead{\AA}  & \colhead{km sec$^{-1}$}  &  \colhead{km sec$^{-1}$} 
 & \colhead{km sec$^{-1}$}   & \colhead{km sec$^{-1}$} \\
}
\startdata
 1321 &   1331  &   -87.7  &   0.5 & 71.0 &    0.4   \\
 1334 &   1338  &      --- &  ---- &  7.2  &   9.6   \\
 1338 &   1348  &  -110.2  &  19.1  &  79.6  &  28.9   \\
 1348 &   1355  &   -73.8  &   1.3 &   84.2  &   0.5   \\
 1355 &   1365  &   -53.2  &   1.8 &   66.4 &    1.9   \\
 1365 &   1371  &   -89.5  &   0.1  &  85.4 &    0.1   \\
 1371 &   1378  &   -93.4  &   5.7 &   86.8  &   3.5   \\
 1378 &   1386  &   -71.9  &   0.7 &   91.7  &   1.5   \\
 1386 &   1395  &   -29.1  &   1.8 &   84.8  &   0.7   \\
 1395  &  1402  &   -79.2  &  14.9 &   83.8 &    2.9   \\
 1402  &  1412 &   -109.0 &   54.7 &   70.1  &  11.0   \\
 1412  &  1418  &   -84.3  &   0.4  &  86.5  &   0.4   \\
 1418  &  1428   &  -76.6  &   0.1 &   70.2 &    0.1   \\
 1428  &  1437  &   -78.1  &   1.0  &  71.7  &   0.8   \\
 1437  &  1449  &   -79.5  &   3.0  &  86.9  &   3.0   \\
 1449  &  1458  &   -92.0  &  12.8 &   83.5  &   7.6   \\
 1458  &  1467  &   -88.2 &    2.4  &  78.2  &   1.4   \\
 1467  &  1472   &  -69.4   &  4.3 &   73.6  &   3.1   \\
 1472  &  1482  &   -73.9  &   2.4  &  81.5  &   0.9   \\
 1482  &  1490  &   -60.9  &    29.3 &   93.7 &   29.8   \\
 1490  &  1498  &   -66.4  &   0.6 &   78.4  &   0.5   \\
 1498  &  1506  &   -90.0  &   2.1  &  80.6  &   1.5   \\
 1506  &  1512  &   -69.6  &  14.1 &   76.8 &   13.5   \\
\enddata
\end{deluxetable}

\begin{figure}
\plotone{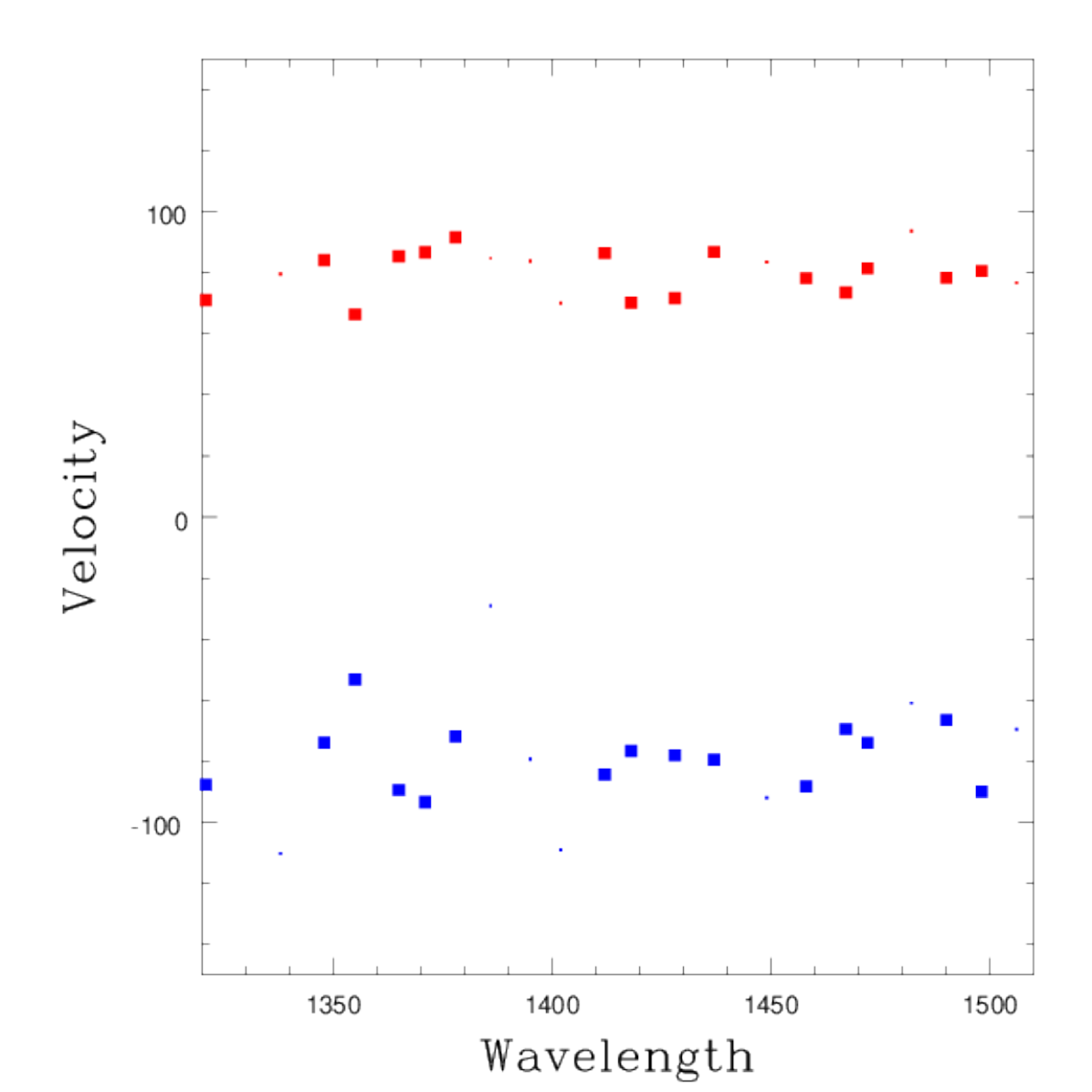}
\caption{Velocities for the two components of the AW Per spectrum. Colors 
correspond to  red and blue velocity components.  Large squares are used in 
the means; dots are velocities which were rejected because of large errors
(see text).  
 Wavelength is in \AA; velocities are in km sec$^{-1}$.   
\label{vr.comp}}
\end{figure}

\begin{figure}
\plotone{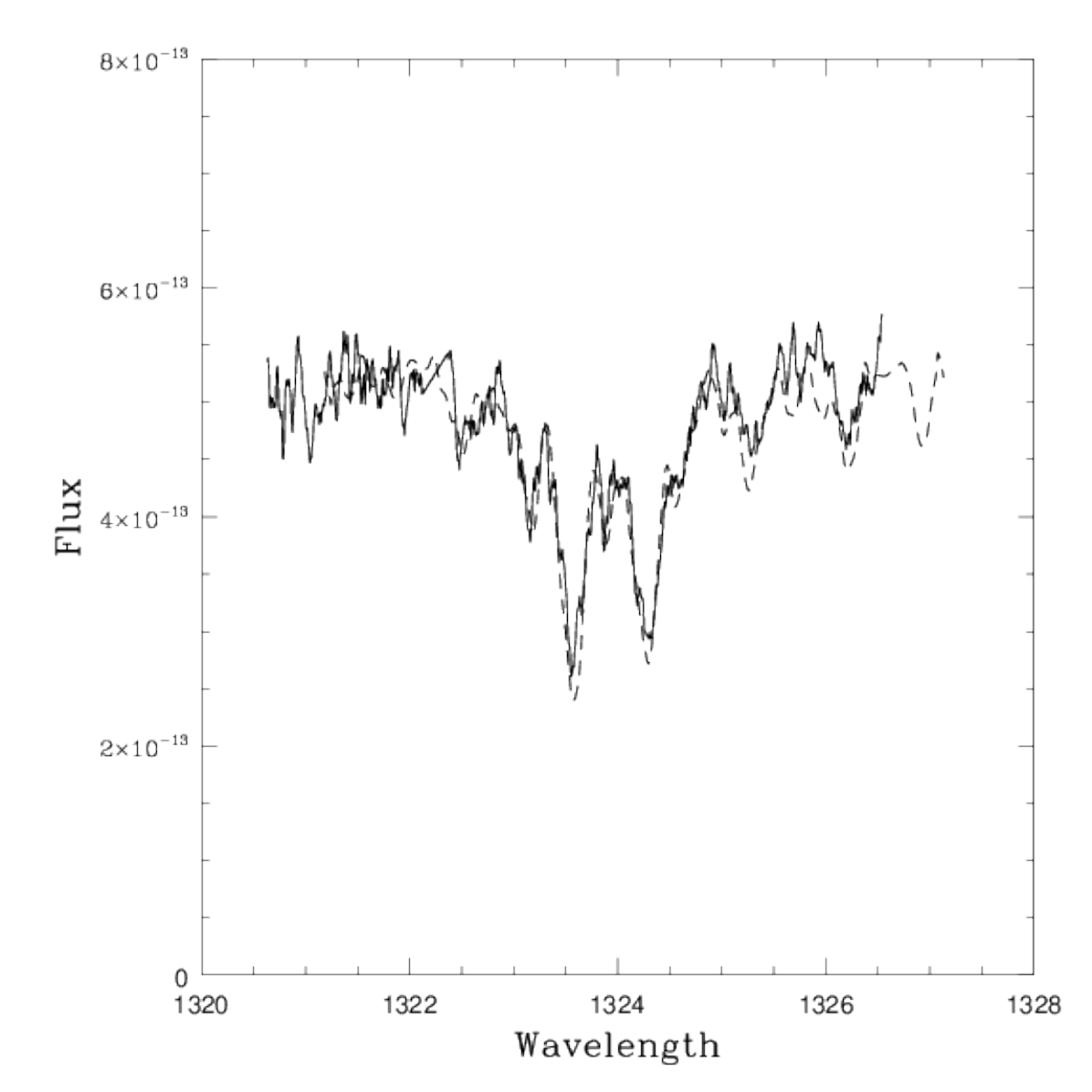}
\caption{The AW Per spectrum (solid line) compared with a simulated
spectrum created from the $\xi$ Oct spectrum (dashed line).  See text for 
details.  The AW Per spectrum   
 has a 10 point boxcar smooth.   Wavelength is in \AA; flux
is in ergs cm$^{-2}$ s$^{-1}$ \AA$^{-1}$.   
\label{comp}}
\end{figure}

Velocities on the coadded STIS spectrum
were measured in the same way as for V350 Sgr using cross-correlation.  In this
case a synthetic spectrum approximating a B7 star with a temperature of 13000 K and a surface 
gravity  log g of 4.0 was used, which responded to both components.  
 It was generated from the synthetic spectrum code SYNTHE (Kurucz 2017) with an ATLAS9 stellar
atmosphere model (Castelli and Kurucz 2003; Kurucz 2017).  

Velocities of both components (B blue and R red  for blue shifted and red shifted components respectively)
for each wavelength region are listed
in Table~\ref{vr} and shown in Fig~\ref{vr.comp}.  
A mean velocity was computed for each component omitting wavelength regions 
where either component had a standard deviation larger than 10 km sec$^{-1}$ (as well as one 
region with a very discordant velocity).  All standard deviations were set to at least 
0.5 km sec$^{-1}$.  The results are   
 -80.6   $\pm$   2.2 and   79.0    $\pm$      1.8  km sec$^{-1}$, with a velocity difference of 
 158.6  $\pm$  2.8.  

A spectrum approximating the two components of AW Per B was created from $\xi$ Oct
adding two spectra separated by the 159 km s$^{-1}$, the velocity separation between
the components.  Because the blue component is slightly deeper than the red 
component, they are designated Ba and Bb respectively.  Four versions of the 
summed spectrum were created with a ratio between the components f(Ba)/f(Bb) of 0.95,
1.05, 1.10, and 1.15.  The best agreement is for a ratio of 1.05, shown in 
Fig~\ref{comp}.  Thus the line depths, and hence presumably the fluxes at 1324 \AA\/
differ by only a small amount.

\subsection{Low Resolution}\label{lores}

AW Per was observed with the {\it IUE} satellite
in low resolution mode (resolution 6 \AA). The spectrum SWP 27541 is discussed here.   The high resolution STIS spectrum discussed in Section~\ref{hires} permits the identification of the two components of the B component (Ba and Bb). In low resolution they are not resolved.  Since Ba and Bb are very similar in temperature and luminosity, we determine a single temperature from the composite spectrum in this section.

\subsubsection{E(B-V)}\label{ebv}
The first consideration in interpreting an ultraviolet spectrum is
determining the reddening.
In particular, for AW Per, early discussions of the {\it IUE} spectrum
(Evans 1989) incorporated both a substantial E(B-V) (0.52 mag) and also a significant
contribution from the companion. We revisit
this topic briefly here. Moffett and Barnes (1985) provide an extensive catalog of Cepheid
observations on the Johnson system, from which Groenewegen (1999) provided a  transformed  I
magnitude on the Cousins system.  

Anticipating the result below, as a first exploration we use a companion temperature of
14000 K, based on E(B-V) = 0.52 (Evans 1994).  The absolute magnitude for the Cepheid
is calculated from the period (6.4636 $^d$) and the period-luminosity law (Leavitt Law) in
V from Cruz Reyes and Anderson (2023), M$_V$ = -3.89 mag.
This is based on {\it Gaia} DR3 parallaxes of
Cepheids and open clusters.  The temperature of the companion corresponds closely
to a B7 main sequence star in the calibration of Pecaut and Mamajek (2013), providing
colors of B-V = -0.128 mag and V-I = -0.133.  This corresponds to an absolute magnitude
M$_V$ = -0.9 mag (Drilling and Landolt 2000), making the companion 2.99 mag fainter
than the Cepheid in V.  The colors of the companion are then reddened to the E(B-V)
of  the Cepheid, becoming B-V = 0.392 mag and (V-I) = 0.517 mag, and from these
B and I of the companion are created (Table~\ref{corr}).
From the $<$B$>$ - $<$V$>$  and 
$<$V$>$ - $<$I$>$ colors, the E(B-V) is calculated using the appropriate
formula from Fernie (1990) for BV(I)$_C$ system to be 0.452 mag
(Table~\ref{red}). 

From the discussion of the HST high resolution spectra in Section~\ref{hires},
the companion is made up of two stars of nearly equal 
temperature and mass.  The effect of two stars of equal temperature
is shown in Table~\ref{corr} below.
The reddenings  from the  models of the companion based on one or two
hot stars are quite similar (Table~\ref{red}).  For comparison, the E(B-V) before
correcting for the effect of the companion is included.  The last line in
Table~\ref{red} is the reddening from Fernie (1990), which is a little different
since it is the average from photometric systems.  Groenewegen (2018) recommends
scaling the Fernie reddenings by 0.94, which is close to 0.50.



\vfill
\eject
\newpage 


\begin{deluxetable}{lrrr}
\tablecaption{Correction for Companion\label{corr}}
\tablewidth{0pt}
\tablehead{
 \colhead{} &  \colhead{$<$ B $> $}  & \colhead{$<$ V$>$}  & \colhead{$<$ I$>$}  \\
}
\startdata
Comp: 1 B7 Star  & & & \\
Comp     &    10.868    &  10.476  &  9.959  \\ 
Cep + Comp &  8.540  &     7.486    & 6.227 \\  
Cep     &     8.675  &     7.557  &   6.260 \\    
Comp: 2 B7 Stars  & & & \\
Comp &       10.118 &      9.726 &   10.243  \\
Cep + Comp &  8.540 &      7.486 &    6.227 \\
Cep    &     8.829  &     7.634 &    6.256 \\
\enddata
\end{deluxetable}

\begin{deluxetable}{lrrr}
\tablecaption{Reddening\label{red}}
\tablewidth{0pt}
\tablehead{
 \colhead{} &  \colhead{$<$ B $>$  - $<$V$>$}  & \colhead{$<$ V$>$ - $<$I$>$}  & \colhead{E(B-V)}  \\
}
\startdata
Comp: 1 B7 star  &   1.118  &  1.297     &    0.452   \\
Comp: 2 B7 stars   &   1.195 &  1.378   &      0.495  \\
Uncorrected  &        1.054    &          1.259   &       0.457  \\
Fernie     &      ---    &           ---      &  0.522  \\
\enddata
\end{deluxetable}









\subsubsection{Companion Temperature}\label{tcomp}

 The {\it IUE} spectrum has been reanalyzed here using Kurucz  ATLAS9
 BOSZ atmospheres (Bohlin, et al. 2017) 
 This was done in the same way as for detached eclipsing binaries (DEBs)
 containing hot stars (Evans. et al. 2023).

\begin{figure}
\plotone{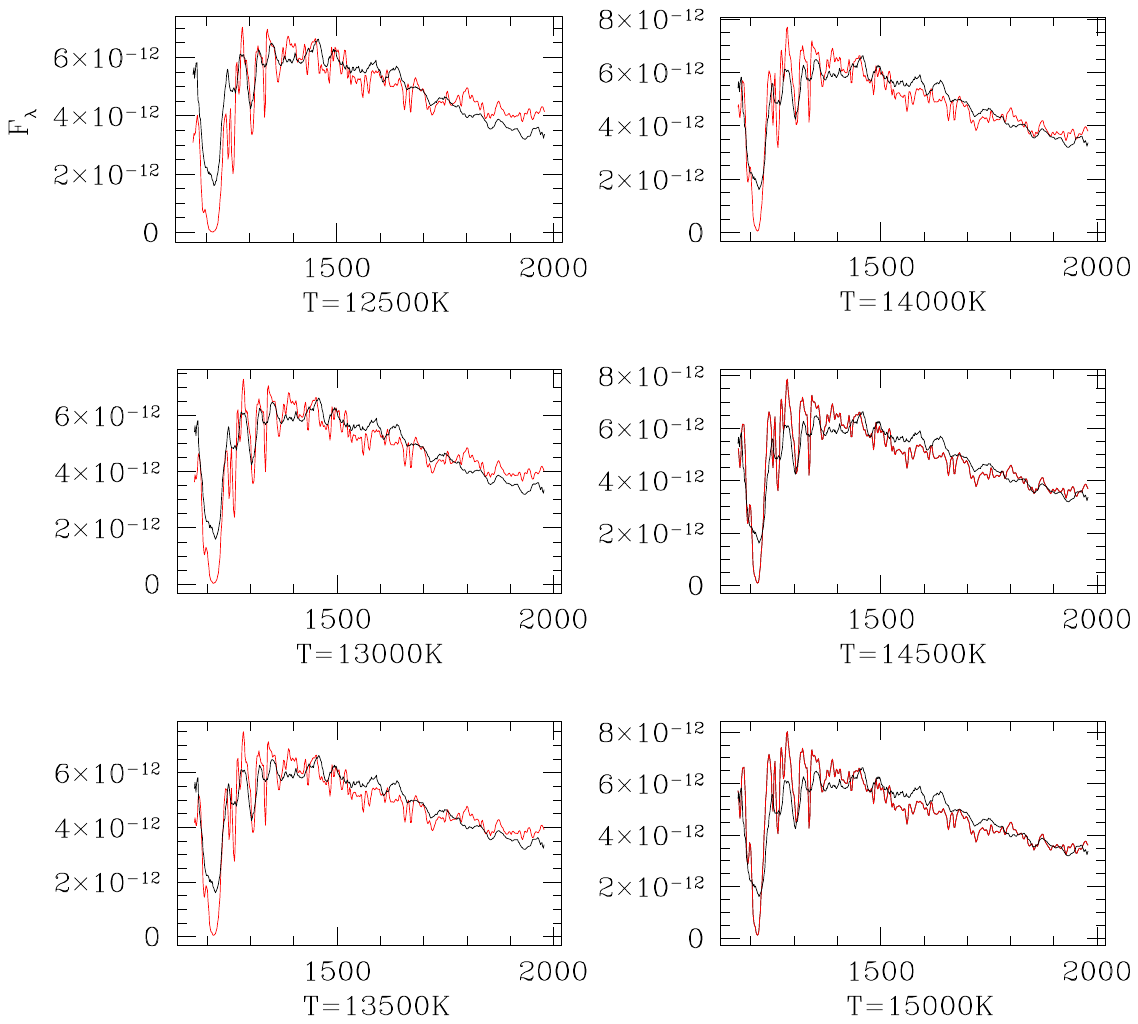}
\caption{The comparison between the BOSZ models (red) and the {\it IUE} spectrum
  (black) dereddened  with E(B-V) = 0.50 mag. The wavelength (x-axis) is in \AA;  flux is in
  ergs cm$^{-2}$ s$^{-1}$ \AA$^{-1}$. The temperature for the models is beneath
  each panel. The region between 1180 and 1250 \AA\/ is shown in the plots, but
  is not included in the temperature determination because of contamination
  from interstellar Ly $\alpha$.  The region between 1250 and 1350 \AA\/ has
  several strong lines or multiplets of Si II and C II.  
\label{spec.model}}
\end{figure}

 The region between 1250 and 1350 \AA\/ has several strong lines from low excitation
 states from Si II and C II, which are apparent in Fig~\ref{spec.model}.
 The Si II lines have been shown to be temperature sensitive (Massa 1989).
 In the low resolution spectrum we have not attempted to model them, but regard them as
 an extra source of uncertainty in the fit.  In AW Per, for example, the lines 
  are made up of two components separated in velocity, making modeling beyond
 the scope of the low resolution spectrum discussion.

\begin{figure}
\plotone{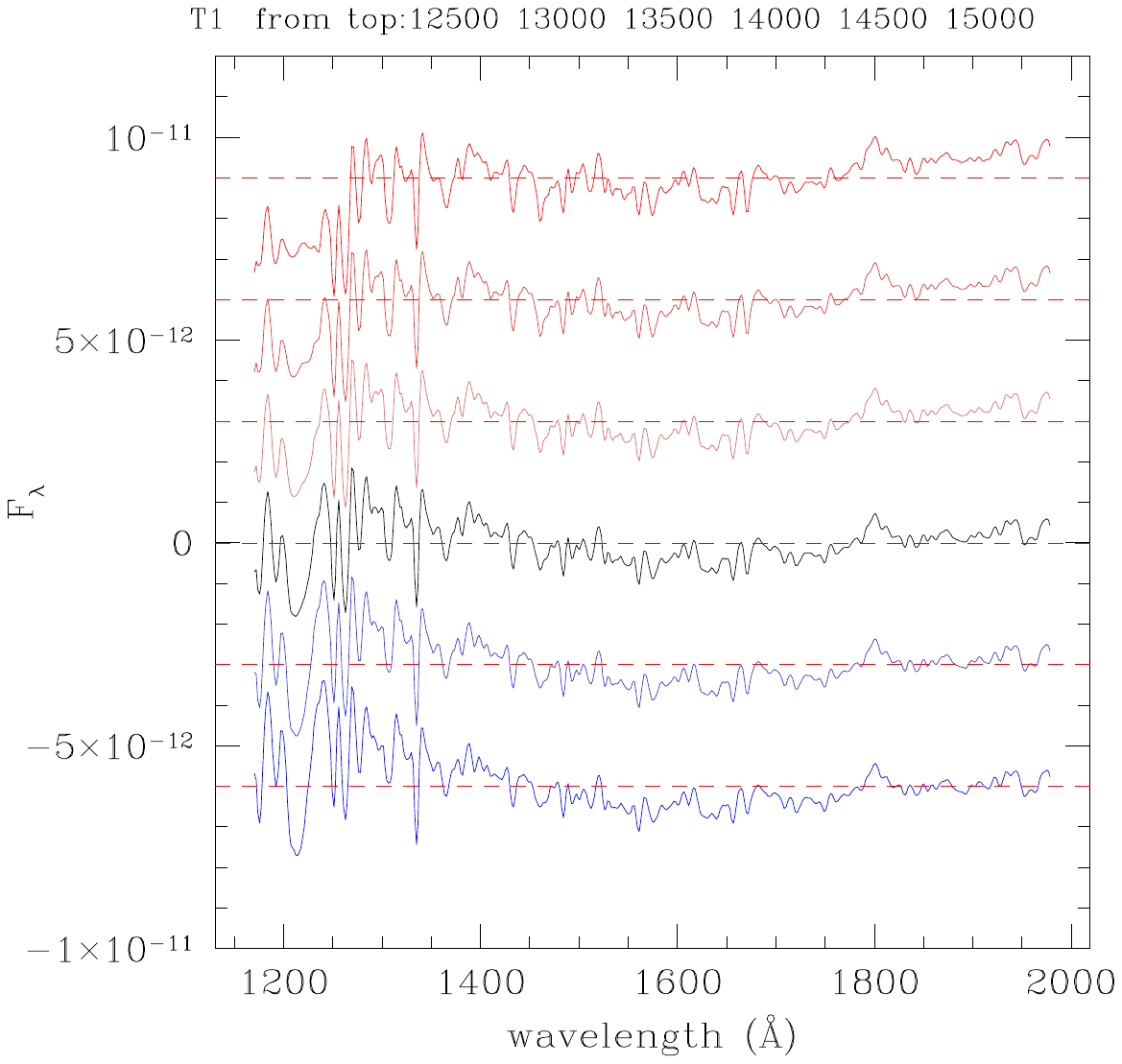}
\caption{The difference between the model and the spectrum [dereddened by E(B-V) = 0.50 mag].  The
  temperatures for the models are listed at the top. The difference spectrum in
  black is the best fit. 
\label{diff}}
\end{figure}

\begin{figure}
\begin{center}
\includegraphics[width=3in]{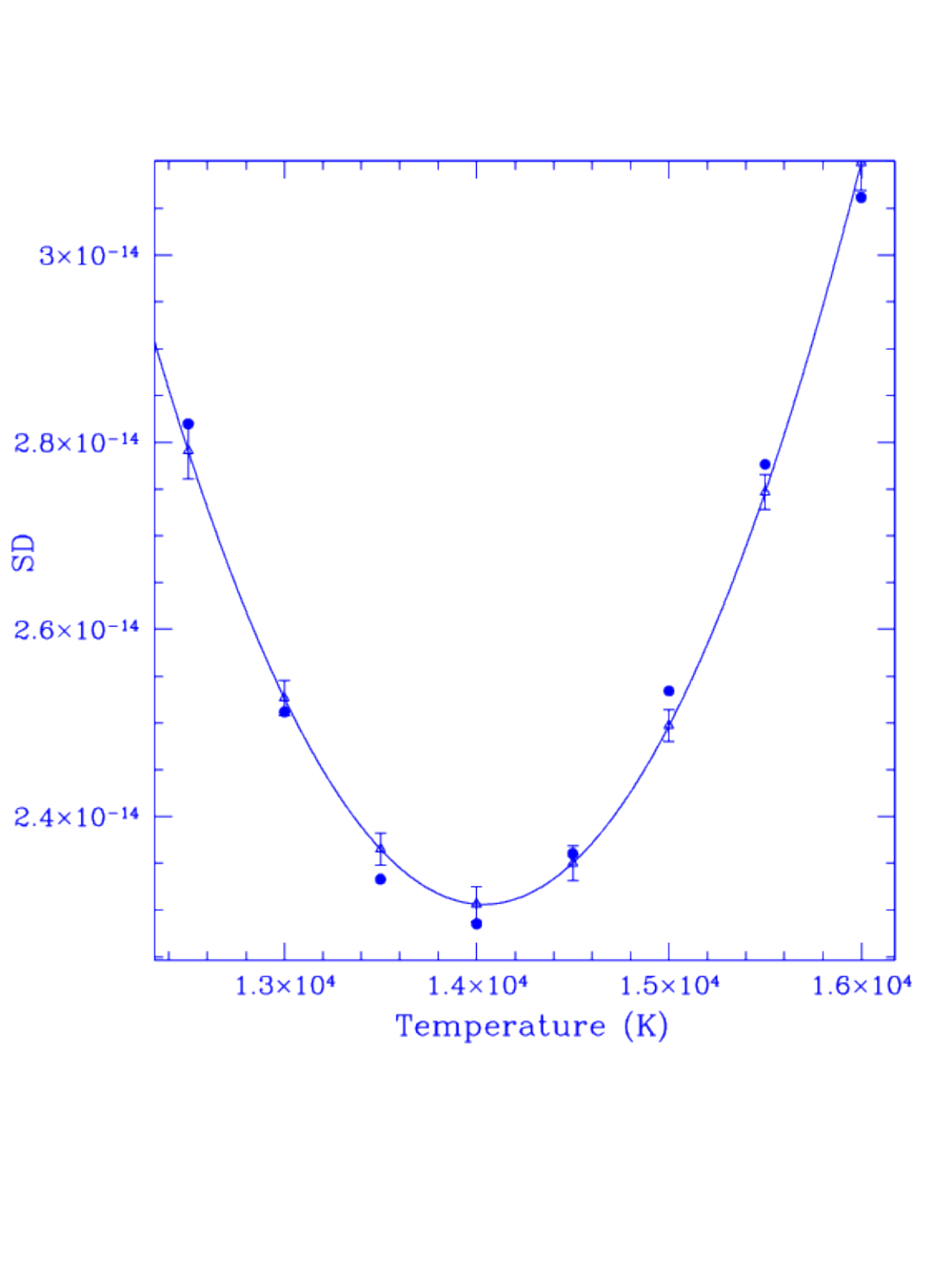}  
\caption{The standard deviations from the spectrum-model comparisons as the
  temperature of the models is changed. Dots: the standard deviation; triangles:
  the parabola fit.  
  \label{para}}
\end{center}
\end{figure}

 The {\it IUE} spectrum was dereddened using E(B-V) = 0.50 mag (Table~\ref{red}).
 The comparisons with the models 
  (spectral fits) are shown in Fig.~\ref{spec.model};  the differences
 between the spectrum and the models are shown in Fig.~\ref{diff}; 
 The  resulting temperature was 14036  $\pm$ 1079 K
 from the parabola fit to the standard deviations of the spectrum-model
 differences (Fig.~\ref{para}).
 As in the discussion
 of the DEBs, the uncertainty was also estimated visually from the spectral
 differences (Fig.~\ref{diff}) to be 500 K.  This temperature is sufficiently close to
 the input temperature used in estimating the effect of the companion on the
 reddening that no further iteration is necessary.  As discussed in Evans, et al. (2023),
 the temperature is only mildly sensitive to E(B-V) because of the
 relatively small wavelength range of the spectrum and the structure in the
 energy distribution.

 The study of the DEBs combined the temperatures with
 their masses  to produce a Mass-Temperatures relation.  Since the two components of the B system Ba and Bb are essentially identical,   the temperature here directly provides  masses for both.
 This is discussed further in Section~\ref{discuss}.


\section{Relative Astrometry}\label{interfer}
	\subsection{CHARA Interferometry}
	\subsubsection{Data acquisition}
	
	Long-baseline optical interferometric data were collected with the Michigan InfraRed Combiner
        (MIRC)[data before 2017.5; Monnier, et al. 2004] and the Michigan InfraRed Combiner-eXeter
        (MIRC-X,)[data after 2017.5; Anugu, et al. 2020], installed at the Center for High
        Angular Resolution Astronomy (CHARA) Array (ten-Brummelaar et al. 2005)
        at the Mount Wilson Observatory.  
        MIRC-X is actually an upgrade of the MIRC instrument with
        mostly an improved sensitivity and wavelength coverage. In addition, recent improvements
        at CHARA include the commissioning of a new 6-telescope beam combiner MYSTIC
        (Michigan Young STar Imager at CHARA; Monnier et al. 2018; Setterholm, et al. 2023), designed alongside the
        MIRC-X upgrade and capable of simultaneous observations. The CHARA array consists of six 1\,m aperture telescopes with an Y-shaped configuration (two telescopes on each branch), oriented to the east (E1, E2), west (W1,W2) and south (S1, S2), and so offering a good coverage of the $(u, v)$ plane. The baselines range from 34\,m to 331\,m, providing an angular resolution down to 0.5 mas  at $\lambda = 1.6\,\mu$m.

	MIRC combined the light coming from all six telescopes in the $H$ band ($\sim 1.6\,\mu$m), with three spectral resolutions (R = 42, 150 and 400). The recombination of six telescopes gives simultaneously 15 fringe visibilities and 20 closure phase measurements that are our primary observables. MIRC-X also combines the light from six telescopes, with the spectral resolution R = 50, 102,  190, and 1170. Our MIRC and MIRC-X observations used only the lowest spectral resolution. MYSTIC is a $K$-band instrument working similarly to MIRC-X and offering a spectral resolutions of 50, 280, 1000, and 1700.
        The log of our observations is available in Table~\ref{table__log}.

	\begin{deluxetable*}{ccccccc}
		\tablecaption{\label{table__log} Log of our MIRC, MIRC-X and MYSTIC observations.}
		\tablewidth{0pt}
		\tablehead{
		\colhead{UT Date}  & \colhead{JD} & \colhead{Inst.}  & \colhead{$N_\mathrm{spec}$} & \colhead{$N_\mathrm{vis}$} & \colhead{$N_\mathrm{CP}$} & \colhead{Calibrators}  }
		\startdata
		2012 Oct. 01   &  2456201.93097 & MIRC &  8 &  1657 &  2232 &  1,2\\
		2013 Sep. 14  &  2456549.96703 &  MIRC & 8 &  1735 & 2267  & 3,4 \\
		2013 Sep. 25 &  2456560.92238 & MIRC & 8  & 561  &  271  & 3,4 \\
		2014 Oct. 01 &  2456931.99035 & MIRC & 8 & 906 & 747 &  5,6 \\
		2015 Oct. 22 & 2457318.00959 & MIRC & 8 & 447 & 238 & 4,5 \\
		2015 Oct. 23 & 2457318.83962 & MIRC & 8 & 334 & 181 & 4,5 \\
		2019 Oct. 15 & 2458771.960780 & MIRC-X & 8 & 1080 & 1440 & 7,8,9 \\
		2021 Sep. 20 & 2459477.955772 & MIRC-X & 8 & 2073 & 2555 & 7,8,9 \\
		2021 Sep. 20 & 2459477.951951 & MYSTIC & 10 & 3378 & 4058 & 7,8,9 \\
		2022 Nov. 19 & 2459902.807354 &MIRC-X & 8 & 3315 & 3021 & 10,11 \\
		\enddata
		\tablecomments{$N_\mathrm{spec}$: number of spectral channel. $N_\mathrm{vis}$: number of visibility measurements. $N_\mathrm{CP}$: number of closure phase measurements. The calibrators used have the following angular diameters: 
			1: $\theta\mathrm{_{LD}(HD19845)} = 0.788\pm0.056$\,mas, 2: $\theta\mathrm{_{LD}(HD30825)} = 0.564\pm0.040$\,mas, 3:  $\theta\mathrm{_{LD}(HD17573)} = 0.414\pm0.029$\,mas, 
			4: $\theta\mathrm{_{LD}(HD25867)} = 0.551\pm0.039$\,mas, 5: $\theta\mathrm{_{LD}(HD29645)} = 0.522\pm0.037$\,mas, 6: $\theta\mathrm{_{LD}(HD30090)} = 0.521\pm0.037$\,mas,	
			7: $\theta\mathrm{_{LD}(HD30099)} = 0.453\pm0.010$\,mas, 8: $\theta\mathrm{_{LD}(HD280306)} =  0.427\pm0.009$\,mas, 9: $\theta\mathrm{_{LD}(BD+38.1014)} =  0.456\pm0.010$\,mas,
			10: $\theta\mathrm{_{LD}(HD30586)} = 0.464\pm0.010$\,mas, 11: $\theta\mathrm{_{LD}(HD276662)} =  0.476\pm0.010$\,mas
		}
	\end{deluxetable*}
	
	We followed a standard observing procedure, i.e., we monitored the interferometric
        transfer function by observing a calibrator before and after our Cepheids. The
        calibrators were selected using the SearchCal
        software\footnote{\url{http://www.jmmc.fr/searchcal}} (Bonneau, et al. 2011)
        provided by the Jean-Marie Mariotti Center. They are listed in Table~\ref{table__log}.
	
\begin{figure*}[ht]
		\centering
		\resizebox{\hsize}{!}{\includegraphics{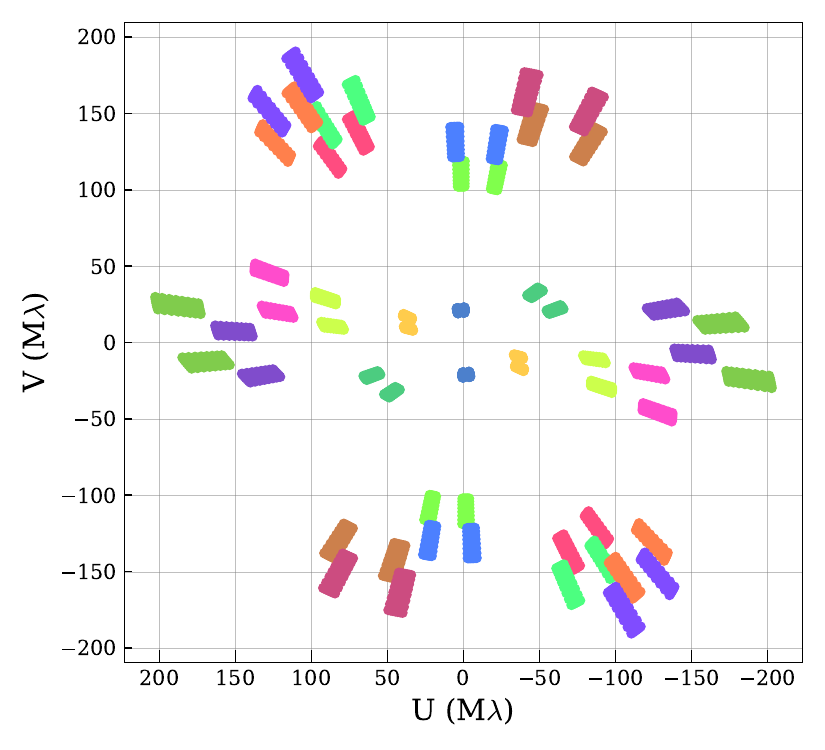}\includegraphics{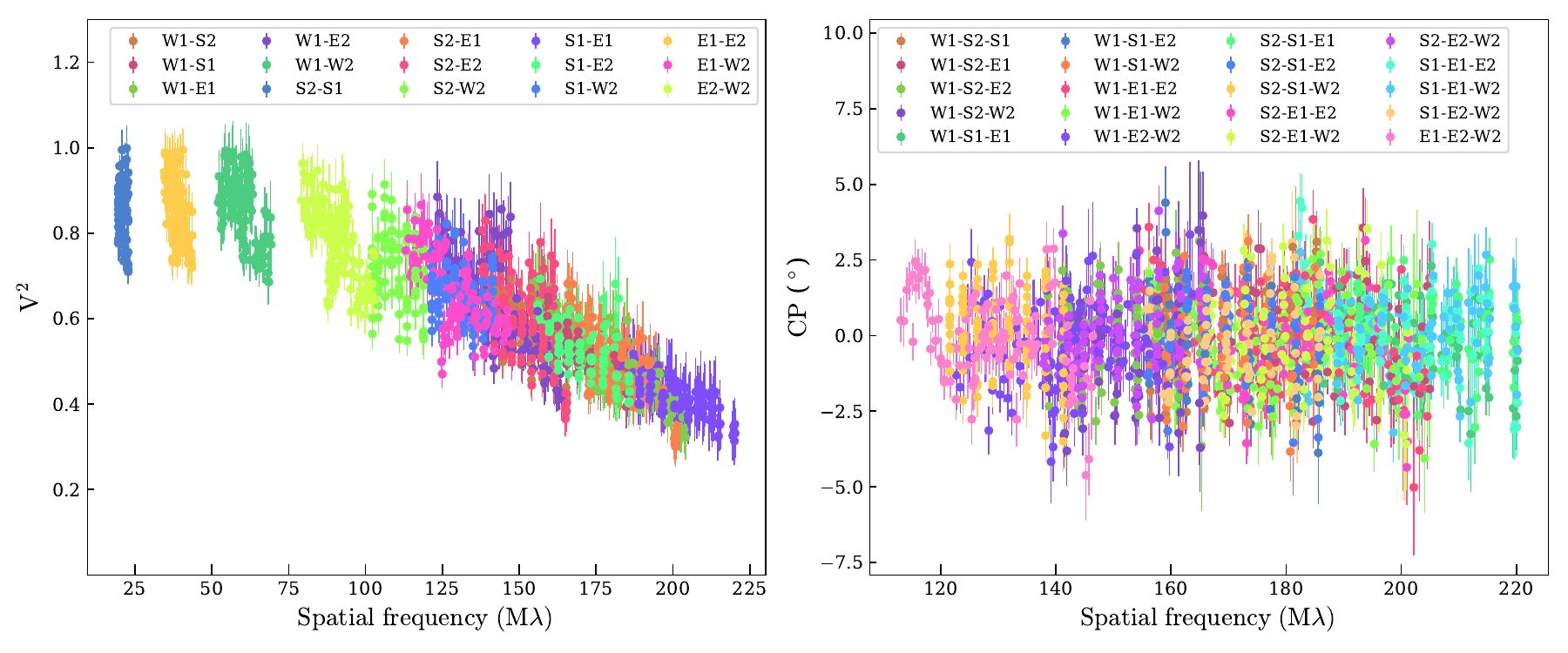}}
		\caption{$(u,v)$ coverage, squared visibilities and closure phases data for the observations of 2012 Oct. 01.}
		\label{figure__data}
	\end{figure*}

	The data were reduced with the standard MIRC and MIRC-X pipelines\footnote{\url{https://gitlab.chara.gsu.edu/lebouquj/mircx_pipeline}}
        (Monnier, et al. 2007;Anugu, et al.  2020). (For 2019 and 2021-2022 the MIRC pipelines
are versions 1.3.3 and 1.4.0 respectively.)        
        The main procedure is to compute squared visibilities and triple products for each baseline and spectral channel, and to correct for photon and readout noises. The data acquired in 2012-10-01 are displayed in Fig.~\ref{figure__data}. MYSTIC data are also reduced with the MIRC-X pipeline.
	
	\subsubsection{Data analysis}
	
	To detect the companion or to be more precise the center of light (in our specific case of a
        binary companion), 
         we used the interferometric tool
        CANDID\footnote{Available at \url{https://github.com/amerand/CANDID}
          and \url{https: //github.com/agallenne/GUIcandid} for a GUI version.}
        (Gallenne, et al. 2015). The main function allows a systematic search for companions performing an $N \times N$ grid of fits, the minimum required grid resolution  which is estimated a posteriori in order to find the global minimum in $\chi^2$. The tool delivers the binary parameters, namely the flux ratio $f$ , and the relative astrometric separation $(\Delta \alpha, \Delta \delta)$, together with the uniform-disk angular diameter $\theta_\mathrm{UD}$ of the primary star (the Cepheid). The angular diameter of the companion is assumed to be unresolved by the interferometer. The significance of the detection is also given, taking into account the reduced $\chi^2$ and the number of degrees
        of freedom \footnote{The computer precision limits the conversion to S/N values greater than 8}
        and comparing it with a single uniform disk model that best fit the data.
        They are listed in Table~\ref{table__astrometry_results}, together with our measured astrometric positions. For the observations of 2015, we combined the dataset in order to increase the detection level. In this case, we only used the closure phase because of the variation of the angular diameter between these two observations ($\sim 10$\,\%, which was fixed to the average fitted value). For Oct. 2014, we also only used the CPs because of a low detection level when including the squared visibilities, likely due to poor seeing conditions which mostly affect this observable.
	
Uncertainties on the fitted parameters are estimated using a bootstrapping function. From the distribution, we took the median value and the maximum value between the 16th and 84th percentiles as uncertainty for the flux ratio and angular diameter. For the fitted astrometric position, the error ellipse is derived from the bootstrap sample (using a principal components analysis). The angular diameters and astrometric positions were then multiplied by factors of $1.004 \pm 0.0025$\,mas for MIRC and divided by $1.0054 \pm 0.0006$\,mas for MIRC-X (J.D. Monnier, private communication) to take into account the uncertainty from the wavelength calibration. This is equivalent to adjusting the respective wavelengths reported in the OIFITS files by the same factors. Uncertainty on the angular diameter measurements was estimated using the conservative formalism of Boffin et al. (2014) as follows:
	\begin{displaymath}
		\sigma^2_\mathrm{\theta_{UD}} = N_\mathrm{sp} \sigma^2_\mathrm{stat} + \delta \lambda^2 \theta^2_\mathrm{\theta_{UD}}
	\end{displaymath}
where $N_\mathrm{sp}$ is the number of spectral channels, $\sigma^2_\mathrm{stat}$ the uncertainty from the bootstrapping and $\delta \lambda = 0.25$\,\% or 0.06\,\%, as mentioned above for MIRC and MIRC-X, respectively.

We measured a mean uniform disk diameter in the $H$ band of $\mathrm{\theta_{UD}} = 0.559\pm0.051$\,mas (the standard deviation is taken as uncertainty), which is in agreement at $0.5\sigma$ with the value estimated by Trahin et al. (2021) from a spectro-photometric analysis. Using the Gaia DR3 parallax  with the Lindegren (2021) correction, 
this provides a linear radius of $R = 58.9 \pm 5.6\,R_\odot$, also in agreement with Trahin et al. (2021). We also estimated an average flux ratio in $H$ of $f_\mathrm{H} = 1.78\pm0.51$\,\%.   We have examined the variation of angular diameter as a function of pulsation phase.  The range is
approximately the $\pm$ 6\% expected for pulsation,  but the phase range from 0.18 to 0.64 is too
small for a definitive conclusion about whether it follows the pattern expected for pulsation.

	

\begin{deluxetable*}{ccccccccc}
		\tablecaption{Relative astrometric position of the AW~Per companion.\label{table__astrometry_results}.}
		\tablewidth{0pt}
		\tablehead{
			\colhead{JD} & \colhead{$\Delta \alpha$}	&  \colhead{$\Delta \delta$} & \colhead{$\sigma_\mathrm{PA}$}	& \colhead{$\sigma_\mathrm{maj}$}	&  \colhead{$\sigma_\mathrm{min}$}	&  \colhead{$f$} &  \colhead{$\theta_\mathrm{UD}$} & \colhead{$n\sigma$}	\\
			\colhead{(Day)} & \colhead{(mas)} & \colhead{(mas)} & \colhead{(deg)} & \colhead{(mas)} & \colhead{(mas)} & \colhead{($\%$)}  & \colhead{(mas)} & 
		}
		\startdata
2456201.931 & 29.751 & 12.579 & -176.2 & 0.076 & 0.033 & $1.59\pm0.09$ & $0.636\pm0.016$ & $> 8$\\
2456549.967 & 31.012 & 10.528 & 159.3 & 0.081 & 0.029 & $1.31\pm0.06$ & $0.512\pm0.036$ & $> 8$ \\
2456560.922 & 31.189 & 10.466 & -162.7 & 0.296 & 0.061 & $1.79\pm0.37$ & $0.520\pm0.067$ & 3.8 \\
2456931.990 & 32.116 & 8.026 & -156.3 & 0.112 & 0.042 & $1.28\pm0.18$ & 0.634 & 7.4 \\
2457318.359 & 32.696 & 5.362 & 146.2 & 0.190 & 0.091 & $2.34\pm0.61$ & 0.583 & 7.5 \\
2458771.961 & 31.386 & -4.714 & -176.2 & 0.021 & 0.006 & $1.21 \pm0.04$ & $0.505\pm0.007$ &  $> 8$  \\
2459477.956 & 29.060 & -9.509 & 157.9 & 0.048 & 0.032 & $2.69\pm0.23$  & $0.588\pm0.019$  & $> 8$ \\
2459477.952 & 29.058 & -9.487 & 54.0 & 0.037 & 0.025 & $3.03\pm0.21$\tablenotemark{a} & $0.622\pm0.028$ & $> 8$ \\
2459902.807 & 27.320 & -12.223 & -145.7 & 0.092 & 0.049 & $2.05\pm0.37$ & $0.531\pm0.025$  & 3.2 \\
		\enddata
		\tablenotetext{a}{Flux ratio for this observation was measured in the $K$ band
                  using MYSTIC.  All other flux ratios are  in the $H$ band.}
		\end{deluxetable*}

We examined the astrometric residuals as a function of time.  There is no sign of additional wobble which might be
caused by the binary secondary.
                
\subsection{HST STIS}\label{hst.stis}

\begin{figure}
\begin{center}
\includegraphics[width=3in]{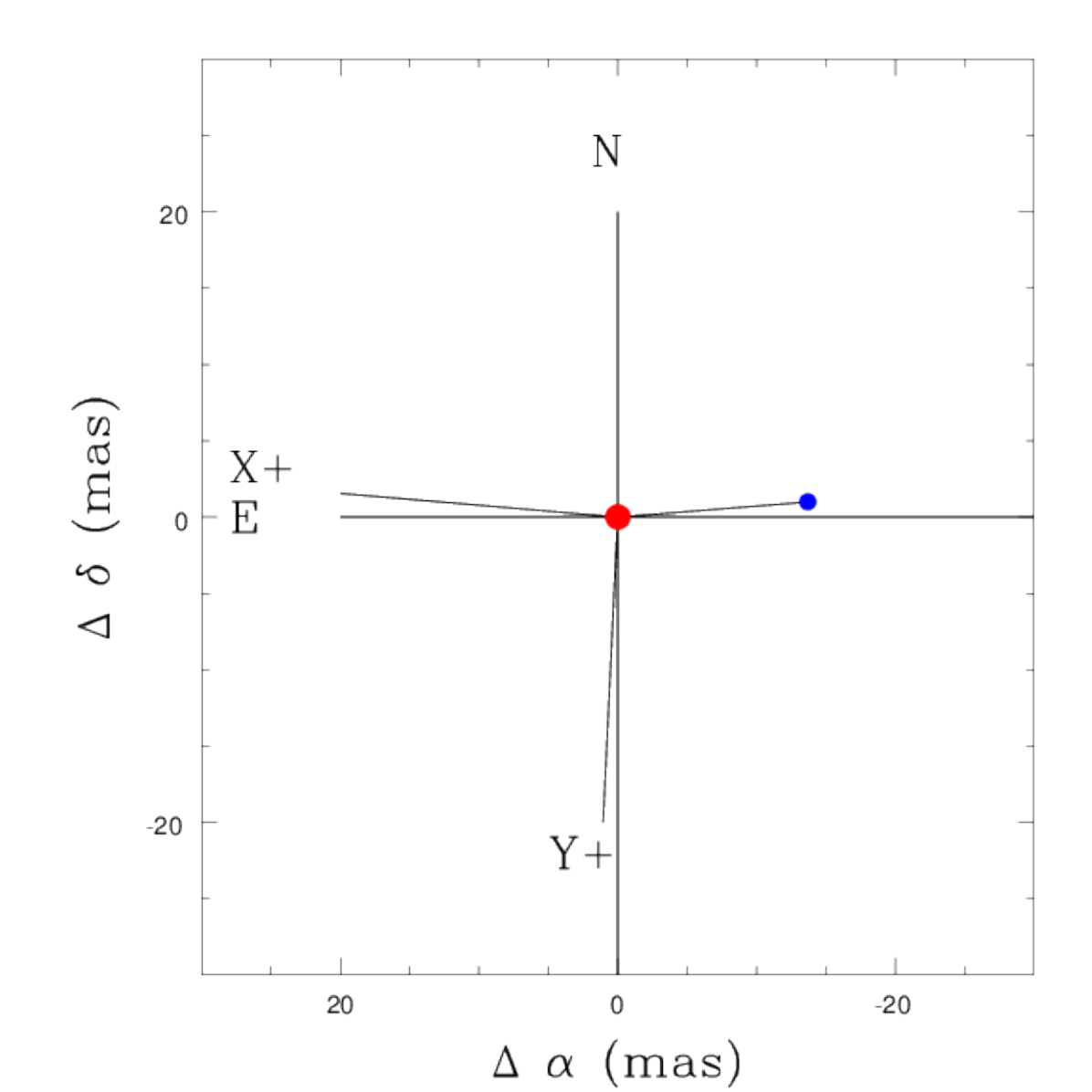}    
\caption{The orientation of the  AW Per A B system.  The figure shows 
the orientation of the STIS field for the roll angle in Massa and Evans of
175.526$^o$, where the X axis is the direction of the dispersion.  The 
red dot is the Cepheid AW Per A and the blue dot is the companion AW Per B 
(itself a binary).   
\label{massa}}
\end{center}
\end{figure}

Massa and Evans (2008) observed the AW Per system with the {\it HST} STIS 
spectrograph. By observing at three spacecraft roll angles they were 
able to derive the separation and angle of components A and B.  It has since
been realized that the roll angle used in the analysis needs to be 
corrected, and the final result is now 13.74 $\pm$ 0.26 mas at a 
position angle of 274.16$^o$ $\pm$ 1.94$^o$ (Fig.~\ref{massa}).

\section{Orbit}\label{orbitt}

The AW Per system now has velocities providing good  coverage of the 38 year orbit.  In 
addition, extensive interferometry now covers more than a quarter of the orbit, 
which provides both the separation between the Cepheid and the companion (itself 
a binary) and the inclination.  The {\it HST} STIS spectra provided the surprise of a
signal from both the hot stars in the companion binary.  This combination
plus a distance provides complete
information about the Cepheid orbit.












\subsection{Pulsation Period}\label{p.puls}

Light curves of Cepheids, and in particular AW Per, have been discussed extensively by Cs\"ornyei, 
et al. (2022).  They have detected the light time in the long-period orbit.  The light-time
effect is $\pm$ 0.05 $^d$.  Over the 15000 days of the Griffin observations, this corresponds
to $\le$ 0.01 in phase, which will not affect the orbital solution. 

Cs\"ornyei, el al. also show 
a parabola in the O-C diagram, presumably an evolutionary change.  
The period change is smaller than the light time effect,  0.113  $\pm$ 0.024 s yr$^{-1}$.
 We have checked to see what the difference in phase is for the span of the 
Griffin observations for the evolutionary period change.  This uses the coefficients of 
the parabola in the O-C diagram (Cs\"ornyei, 2023 private communication).  The difference in 
phase between a constant period and a changing period is 0.0116.  This corresponds to 
$\le$ 0.5 km s$^{-1}$ in pulsation velocity, often significantly less, which adds very little
to the uncertainties in the Griffin data.  For this reason we use a constant period in 
fitting the pulsation velocity curve.  

The period found in the model fitting
solution below 6.463635 $\pm$ 0.000006 $^d$  (using the Epoch from Griffin)
 corresponds to  $\le$ 0.01 difference in phase from the period of Cs\"ornyei   
 over the span of the Griffin observations.  









	\subsection{Model fitting}
	\label{section__model_fitting}

	Orbital parameters are determined by simultaneously fitting the radial velocities (RVs) of the primary star (the Cepheid, which is a single-line spectroscopic binary) and the astrometric positions using a Markov chain Monte Carlo (MCMC, 100\,000 samples with uniform priors)
        routine\footnote{With the Python package \emph{emcee} developed
          by Foreman-Mackey et al. (2013).}. Our RV models include the pulsation of the Cepheid and its orbital reflex motion due to the presence of a companion, while the astrometric model defines the relative astrometric motion of the companion around the Cepheid. Our fitting procedure is
        detailed in Gallenne et al. (2018), who used a linear parametrization technique to solve for the orbital and pulsation parameters. Briefly, the radial velocities are defined by the combination of the pulsation and orbital motion with:
		\begin{equation}
			V(t) = C_1 \cos \nu + C_2 \sin \nu + v_0 + \sum_{i=1}^6 [A_i \cos(2\pi i \phi_\mathrm{puls}) + B_i \sin(2\pi i \phi_\mathrm{puls})]
	\end{equation}
	with $\nu$ the true anomaly of the companion, the pulsation phase $\phi_\mathrm{puls} = (t - T_0)/P_\mathrm{puls}$ (modulo 1) related to the pulsation period and the time of periastron passage, and $(A_i, B_i)$ the amplitude of the Fourier series. The parameters $C_1, C_2$ and $v_0$ are related to the Keplerian elements through the relations of Wright and Howard (2009):
	\begin{eqnarray*}
	K_1 &=& \sqrt{C_1^2 + C_2^2}, \\
	\tan w &=& \frac{-C_2}{C_1},\\
	v_\gamma &=& v_0 - K_1 e \cos \omega.
\end{eqnarray*} 
	with $K_1$ the semi-amplitude of the Cepheid’s orbit due to the companion, $\omega$ the argument of periastron of the companion’s orbit, $e$ the eccentricity of the orbit, and $v_\gamma$ the systemic velocity.


	The astrometric positions of the companion as measured from interferometry are modelled with the following equations:
	\begin{eqnarray}
		\Delta \alpha &=& r \left[ \cos \Omega \cos(\omega + \nu) - \sin \Omega \sin (\omega + \nu) \cos i \right], \\
		\Delta \delta &=& r \left[ \sin \Omega \cos(\omega + \nu) + \cos \Omega \sin (\omega + \nu) \cos i \right],
	\end{eqnarray}
	with $a$ the angular semi-major axis in arcsecond, $\Omega$ the position angle of the ascending node, and $i$ the orbital inclination. The true anomaly $\nu$ and the separation $r$ at a given time $t$ are calculated as:
	\begin{eqnarray*}
		r &=& a (1 - e \cos E), \\
		\tan \dfrac{\nu}{2} &=& \sqrt{\dfrac{1+e}{1-e}} \tan \dfrac{E}{2},
	\end{eqnarray*}
	where eccentric anomaly $E$ is calculated by solving the Kepler's equation $2\pi (t - T_\mathrm{p})/P_\mathrm{orb} = E - e \sin E$.

	We used the RVs from Griffin (2016) which span the full orbital period of 40 years. We did not add additional data from the literature to avoid possible systematics due to zero points offsets.
	
\begin{figure*}[ht]
		\centering
		\resizebox{\hsize}{!}{\includegraphics{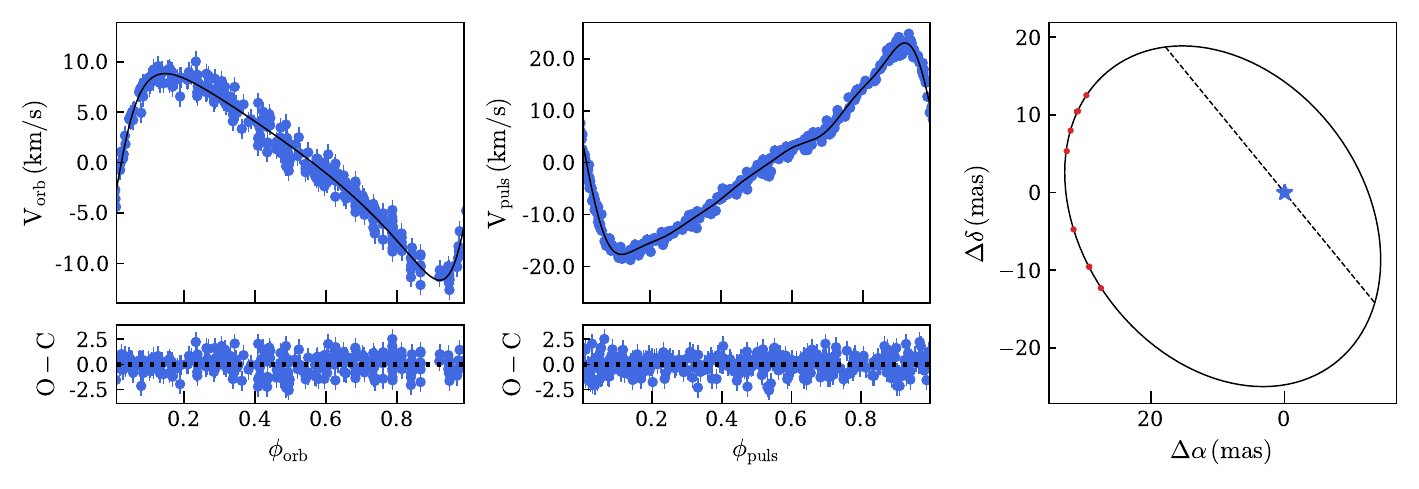}}
		\caption{Result of our combined fit. Left: fitted (solid lines) and measured (blue dots) Cepheid's orbital velocity. Middle: fitted and measured pulsation velocity. Right: relative astrometric orbit of AW~Per Ab. Red points from CHARA.}
		\label{figure__orbit}
	\end{figure*}

	AW~Per is a single-line spectroscopic binary, therefore, the masses and the distance are degenerate parameters. However, we can assume the distance and derive the individual masses (or in our case the
        primary mass and the total mass of the binary companion).
        We took the parallax measured by Gaia DR3 
including the Lindegren correction, 
         $\varpi = 1.0206\pm0.0287$\ mas, which we implemented in our MCMC analysis with a normal distribution centered on 1.0206\,mas with a standard deviation of 0.0287\,mas. Masses are then derived from the distribution with:
		\begin{eqnarray*}
		M_\mathrm{T} &=& \dfrac{a^3}{P_\mathrm{orb}^2\,  \varpi^3} , \\
		q &=& \left[ \dfrac{a \sin{i}}{0.03357\, \varpi\, K_1\, P_\mathrm{orb} \sqrt{1 - e^2}} - 1 \right]^{-1} \\
		K_2 &=& \dfrac{K_1}{q} \\
		M_1 &=& \dfrac{M_\mathrm{T}}{1+q}\\
		M_2 &=& q\,M_1
	\end{eqnarray*}
	        with $M_\mathrm{T} = M_1 + M_2$ the total mass in $M_\odot$, $a$ and $\varpi$ in mas, $P_\mathrm{orb}$ in years, $K_1$ in km\,s$^{-1}$ and $q = M_2/M_1$ the mass ratio. Our orbital fit is displayed in Fig.~\ref{figure__orbit} and the final parameters are listed in Table~\ref{table__results}. Our fitted spectroscopic orbital elements are in very good agreement with those found by Griffin.

	\begin{table}[!h]
		\centering
		\caption{Final estimated parameters of the AW~Per system. Index 1 designates the Cepheid and index 2 its companion.}
		\begin{tabular}{cc|cc} 
		\hline
		\hline
		\multicolumn{2}{c|}{Pulsation} & \multicolumn{2}{c}{Orbit} \\
		\hline
		$P_\mathrm{puls}$ (days)			&   $6.463635 \pm 0.000008$ &   $P_\mathrm{orb}$ (days)									 &  	 $13849.44 \pm  0.83 $ 	  \\
		$T_0$ (JD)						   	&   2451873.616\tablenotemark{a}		&   $T_\mathrm{p}$ (JD)										 & $ 2438763.53  \pm 1.21 $  \\
		$A_1$ (km~s$^{-1}$)					&	 $3.62 \pm 0.02$			&   $e$																     &   $0.465 \pm 0.002$  	\\
		$B_1$ (km~s$^{-1}$)					&	 $-14.93 \pm 0.01$		&   $\omega$	($^\circ$)								  &	$ 251.73 \pm 0.05 $ 		  \\
		$A_2$ (km~s$^{-1}$)					&	 $2.23 \pm 0.01$	&   $K_1$ ($\mathrm{km~s^{-1}}$)					  	&	$10.260 \pm 0.023$ 	\\
		$B_2$ (km~s$^{-1}$)					&	 $-7.09 \pm 0.01$		&   $v_\gamma$	($\mathrm{km~s^{-1}}$)			&	$  7.937 \pm 0.013 $		\\
		$A_3$ (km~s$^{-1}$)					&	 $0.51 \pm 0.01$			&   $\Omega$	($^\circ$)									 &	$44.58\pm 0.08$  	\\
		$B_3$ (km~s$^{-1}$)					&	 $-3.80 \pm 0.01$		&   $i$ ($^\circ$)													&	$46.64 \pm 0.12$ 	\\
		$A_4$ (km~s$^{-1}$)					&	 $0.49 \pm 0.01$		&   $a$ (mas)														 &	$28.781 \pm 0.043$ 	\\
		$B_4$ (km~s$^{-1}$)					&	 $-1.75 \pm 0.01$		&   $a$ (au)														 	&	$ 28.188 \pm 0.802 $ 	 \\
		$A_5$ (km~s$^{-1}$)					&	 $0.40 \pm 0.01$			&   $\varpi$ (mas)												&	$1.021 \pm 0.029$\tablenotemark{b}	\\
		$B_5$ (km~s$^{-1}$)					&	 $-1.02 \pm 0.01$		&   $q$ 																&	$ 1.29 \pm 0.09$   	\\
		$A_6$ (km~s$^{-1}$)					&	 $0.03 \pm 0.01$		&   $M_1$ ($M_\odot$)												&	$6.79 \pm 0.85$ 	\\
		$B_6$ (km~s$^{-1}$)					&	 $-0.51 \pm 0.01$		&   $M_2$ ($M_\odot$)											&	$8.79 \pm 0.50$   	\\
		\hline															
		\end{tabular}
		\tablenotetext{a}{Kept fixed to the value given in Griffin (2016).}
		\tablenotetext{b}{Taken from Gaia DR3 1.0566 $\pm$0.0287mas plus Lindegren correction -0.036038mas.}
		\label{table__results}
	\end{table}

\section{Discussion}\label{discuss}

Several sources of uncertainty contribute to the error in the derived masses, which
are discussed here.

The distance used in Table~\ref{table__results} is derived
from {\it Gaia} DR3.  This needs to be
revisited when the final {\it Gaia} processing is available.
However for such a long orbital period the
results are not as uncertain as for shorter periods. In addition, Khan, et al. (2023)
have discussed parallax uncertainties for comparatively bright stars (G $\simeq$7 mag), which
could add $~$3\% to Gaia parallaxes.  
To give a sense of the uncertainty in masses due to the uncertainty in distance at the
present time, a comparison is provided by the Leavitt Law
(Period-Luminosity relation) from Cruz Reyes and Anderson (2023),
which is based on Cepheids in clusters as
well as parallaxes, all from {\it Gaia} DR3. (The Leavitt Law is used in V so that observed
magnitudes can be corrected for the companion.) The M$_V$ for AW Per from this is -3.893 mag,
corresponding to a distance of 989 pc.
A simple substitution of this distance in Kepler's Third Law yields the sum of the
masses of 15.53 $M_\odot$ instead of 15.58 $M_\odot$ from the solution in
Table~\ref{table__results}.

  The revised astrometry from {\it HST} (Fig.~\ref{massa}) is now in
general agreement with the CHARA astrometry (Fig.~\ref{figure__orbit}).  However, the
uncertainty on the {\it HST} astrometry is larger than that of the CHARA astrometry.
Since that point degrades the solution it was omitted in the solution in
Table~\ref{table__results}.  Inclusion of the {\it HST} results in a Cepheid mass of
5.75 $\pm$ 0.77$M_\odot$.

The combination of the spectroscopic and astrometric orbits allows the determination of 
the  mass of the Cepheid (primary: 6.79 $\pm$ 0.85 $M_{\odot}$) and 
the combined mass of the stars in the binary secondary (8.79 $\pm$ 0.50 $M_{\odot}$).


The coming Gaia DR4 (expected within 2 years) will include an improved parallax
value and absolute epoch astrometry of the displacement of the photocenter of
the AW Per AB system over a period of $\approx 6$\,years. In combination with
the interferometric orbit, the Gaia astrometry will provide important new
constraints on the orbital parameters and total mass of the system.

From Section~\ref{tcomp}, the temperature of the companion AW Per B (the composite Ba +Bb)
is 14036 $\pm$ 500 K.
 Previous temperature determinations using the same {\it IUE} spectrum range from 15739 K
 (Massa and Evans 2008) and 11481 K (Evans 1994) from comparisons with {\it IUE} spectra
 of spectral standards, giving rise to a range of masses.  The temperature found here solidly in
 the middle.


 The Mass-Temperature relation derived from DEBs (Evans, et al. 2023)
 can be used to estimate the masses of  the components of the B system which are treated as essentially identicalbf.  Since Cepheid
 companions must be relatively young, they will be closer to the Zero Age Main
 Sequence than the average DEB and the Mass-Temperature relation is appropriately adjusted
 to be 0.02 smaller in log M than that for DEBs. 
 The mass of  each component Ba and Bb   is thus 3.5 $M_{\odot}$ ranging from 3.3 to 3.7 $M_{\odot}$
 for an uncertainty in T of 500 K.

The STIS echelle spectrum clearly shows the presence of both components (Fig~\ref{vr.comp}), 
 which provides unusually complete information for a triple system.
The relative line strengths in Fig~\ref{lines} provide an estimate of approximately 5\% 
difference  in flux.  It is thus appropriate to estimate the mass of each component of AW Per B
to be half that of M$_2$ in Table~\ref{table__results} (4.4 $\pm$ 0.5  $M_{\odot}$). The
mass inferred from the temperature is close to this range.

A little more can be inferred about the properties of the AW Per B (Ba + Bb) binary 
system. The date of the STIS observations (2,457,302) corresponds to a phase of 0.340 in 
the long period orbit (Table~\ref{table__results}).  Using the mass ratio between the Cepheid and 
the AW Per B binary (q = 1.3), the velocity of the center of mass of the AW Per B system is 
-3.8 km $s^{-1}$, which is essentially zero in Fig.~\ref{vr.comp}. Thus,  the velocity 
difference between components in Fig.~\ref{vr.comp} $\pm$ 79 km $s^{-1}$ is approximately
the orbital velocity.  This velocity is consistent with a period approaching a year for
the two masses in the short period orbit.  Such an orbit would have a semi-major axis
approaching 2 au.  
This is, of course, a lower limit to the orbital velocity since it is from one random 
phase.  In addition the inclination is unknown.   A larger orbital velocity would be 
consistent with a shorter period and a smaller semi-major axis.  Because the orbit 
with the Cepheid  is so long, it can easily accommodate this orbit for the AW Per B.

\begin{figure}
\begin{center}
 \includegraphics[width=4in]{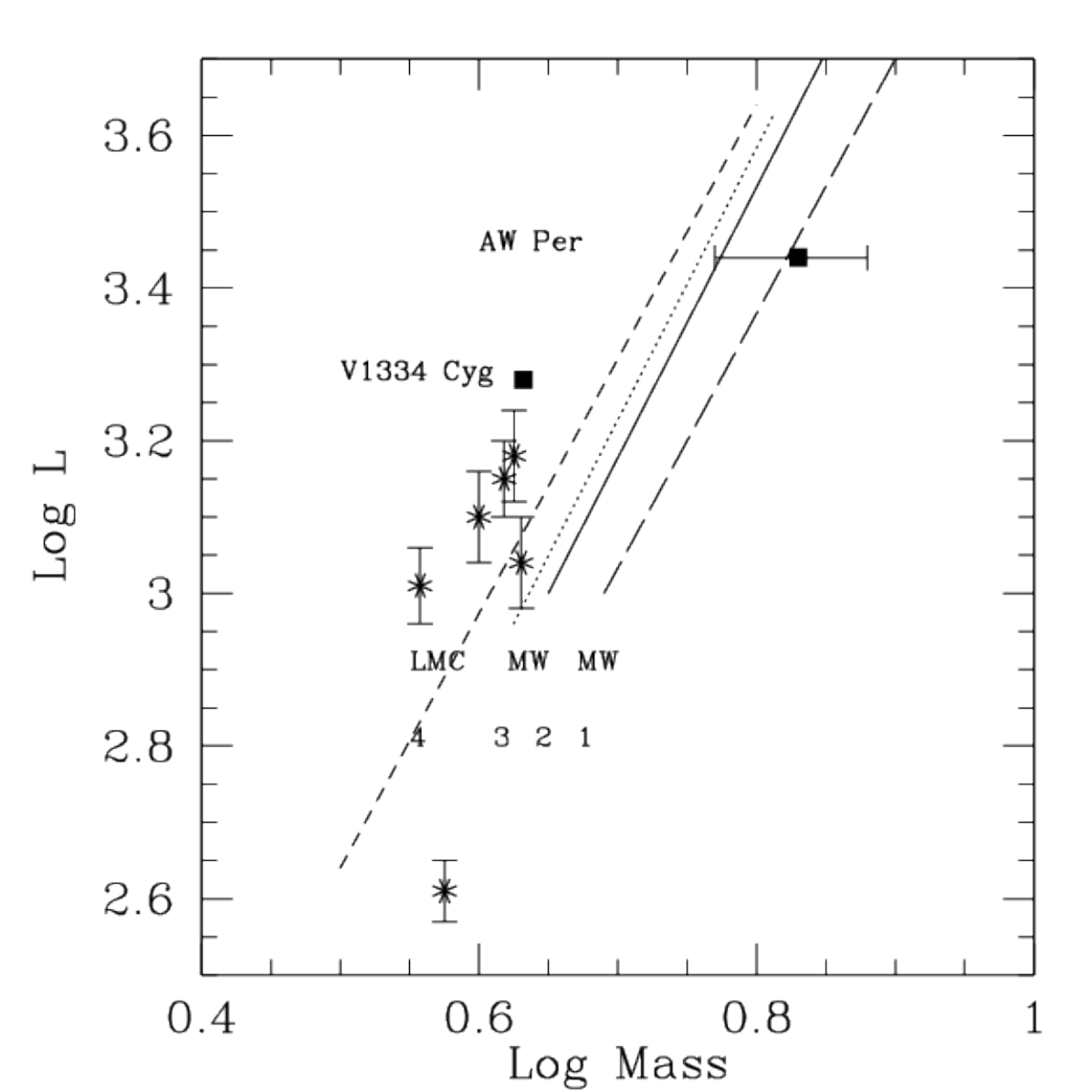}          
\caption{The mass-luminosity relation for Cepheids. Milky Way Cepheids: AW Per
  and  V1334 Cyg; LMC Cepheids: asterisks; predictions from evolutionary tracks:
  long dash (1):  MW metallicity, no main sequence convective overshoot;
  solid:  (2) MW metallicity,  moderate convective overshoot; dotted line: (3)
 MW metallicity,  small convective overshoot and rotation;
 short dash line: (4) LMC metallicity, moderate convective overshoot. Masses and
 luminosities are in solar units.
\label{ml}}
\end{center}
\end{figure}

Fig.~\ref{ml} shows the mass of AW Per in comparison with other masses  for Cepheids from
the Milky Way and the LMC.   The luminosity for AW Per comes from the {\it Gaia} distance as used in Table~\ref{table__results} and the Cepheid magnitude and reddening from Tables~\ref{corr} and~\ref{red}.  Uncertainty in the luminosity will be evaluated further after the {\it Gaia} DR4 release. 
In addition to AW Per, the other Milky Way Cepheid is
V1334 Cyg (Gallenne, et al. 2018), the most accurately determined Cepheid mass
in the Milky Way.  
LMC Cepheids are from Pilecki et al. (2021). The
predictions from evolutionary tracks cover a range of parameters:
1.  MW metallicity, no main sequence convective overshoot (Bono et al. 2016);
2. MW metallicity and moderate core convective overshoot (Bono et al. 2016);
3. MW metallicity, small convective overshoot and rotation (Anderson
et al. 2016);
4. LMC metallicity and moderate core convective overshoot (Bono et al. 2016).
The error bars on the mass of AW Per are too large for it
to be a definitive diagnostic between these parameters.  This
approach is preliminary until the final {\it Gaia} data release provides
a definitive distance.  

\section{Conclusions}\label{}

In summary, the astrometry and velocity coverage of the long period orbit
provide a  mass of the Cepheid AW Per.  The accuracy of mass of the Cepheid
will improve with a distance from the final processing of {\it Gaia} data. The
stars in the binary secondary AW Per Ba and Bb are well separated in velocity
in the HST STIS spectrum, which shows that they
are very similar in mass.  


\section{Acknowledgments}\label{ackn}

It is a pleasure to thank G. Cs\"orynei providing evolutionary period change
from the O-C light curve data
and K. Frey for providing
digital OCR versions of the Griffin velocities.  D. Massa provided instructive
conversations about the {\it HST} STIS astrometry.
Support for EMW was provided from HST-GO-14194.001-A; support for CP and JK was
provided from HST-GO-15861.001-A.
HMG was supported through grant HST-GO-15861.005-A from the STScI under NASA contract NAS5-26555.
Support was provided to NRE by the Chandra X-ray Center NASA Contract NAS8-03060.

This work is based upon observations obtained with the Georgia State University Center for High Angular Resolution Astronomy Array at Mount Wilson Observatory.  The CHARA Array is supported by the National Science Foundation under Grant No. AST-1636624 and AST-2034336.  Institutional support has been provided from the GSU College of Arts and Sciences and the GSU Office of the Vice President for Research and Economic Development.
Time at the CHARA Array was granted through the NOIRLab community access program (NOIRLab PropID: 2019A-0071; PI: A. Gallenne). 
SK acknowledges funding for MIRC-X received funding from the European Research Council (ERC) under the European Union's Horizon 2020 research and innovation programme (Starting Grant No. 639889 and Consolidated Grant No. 101003096). JDM acknowledges funding for the development of MIRC-X (NASA-XRP NNX16AD43G, NSF-AST 1909165) and MYSTIC (NSF-ATI 1506540, NSF-AST 1909165).

 PK, GP, and WG acknowledge funding
from the European Research Council (ERC) under the European Union's Horizon 2020 research and innovation program (projects CepBin, grant agreement No 695099, and UniverScale, grant agreement No 951549).
AG acknowledges financial support from ANID-ALMA fund No. ASTRO20-0059.
and the Agencia Nacional de Investigaci\'on Cient\'ifica y Desarrollo (ANID)
through the FONDECYT Regular grant 1241073.
RIA acknowledges support from the European Research Council (ERC) under the European Union's Horizon 2020 research and innovation programme (Grant Agreement No. 947660). RIA is funded by the Swiss National Science Foundation through an Eccellenza Professorial Fellowship (award PCEFP2\_194638).
RMR acknowledges
funding from the Heising-Simons Foundation 51 Pegasi b Fellowship.

This work has made use of data from the European Space Agency (ESA) mission Gaia (https://www.cosmos.esa.int/gaia), processed by the Gaia Data Processing and Analysis Consortium (DPAC, https://www.cosmos.esa.int/web/gaia/dpac/consortium). Funding for the DPAC has been provided by national institutions, in particular the institutions participating in the Gaia Multilateral Agreement.
The SIMBAD database, and NASA’s Astrophysics Data System Bibliographic Services
were used in the preparation of this paper.

The data presented in this article were obtained from the Mikulski Archive for Space Telescopes (MAST) at the Space Telescope Science Institute. The specific observations analyzed can be accessed
via \dataset[10.17909/9gjn-gx79]{https://doi.org/DOI}




\end{document}